\documentclass[letter]{aa}  %
\usepackage[english]{babel}  
\usepackage{graphicx}  
\usepackage[utf8]{inputenc}  
\usepackage{microtype}  
\usepackage{booktabs}  
\usepackage{rotating}  
\usepackage{lscape}
\usepackage[squaren]{SIunits}
\usepackage{natbib}
\usepackage{amsmath}
\usepackage{amssymb}
\usepackage{pstricks}
\usepackage{txfonts}
\usepackage{tikz}
\usepackage{color}
\usepackage{hyperref} %

\newcommand{\td}{T_\mathrm{{d}}}

\newcommand{\pot}[1]{10^{#1}}

\newcommand{\vlsr}{V_\mathrm{{LSR}}}

\newcommand{\cm}{\usk\centi \metre}

\newcommand{\kpc}{\usk\mathrm{kpc}}
\newcommand{\kpctab}{\mathrm{kpc}}
\newcommand{\hii}{H\textsc{ii}}
\newcommand{\hi}{H\textsc{i}}
\newcommand{\msun}{\usk\mathrm{M_\odot}} 
\newcommand{\msuntab}{\mathrm{M_\odot}}

\newcommand{\jybeam}{\ensuremath{\usk\mathrm{Jy}\usk\text{beam}^{-1}}}
\newcommand{\yr}{\usk\mathrm{yr}}

\newcommand{\lsun}{\usk\mathrm{L_\odot}}
\newcommand{\lsuntab}{\mathrm{L_\odot}}

\newcommand{\kms}{\usk\kilo\metre\usk\second^{-1}}
\newcommand{\kmstab}{\kilo\metre\usk\second^{-1}}
\newcommand{\kel}{\usk\kelvin}
\newcommand{\mum}{\usk\micro\metre}

\newcommand{\asec}{\ensuremath{^{\prime\prime}}}
\newcommand{\unc}[2]{_{#1}^{#2}}
\newcommand{\mvir}{M_\mathrm{vir}}

\newcommand{\nobrkdash}[1]{#1\mbox{-}}

\newcommand{\reff}{\ensuremath{R_\mathrm{eff}}}

\newcommand{\firdconezero}{G010.4}
\newcommand{\firdconeeight}{G018.7}
\newcommand{\firdcthreethreesix}{G336.9}
\newcommand{\firdcthreethreesevenpone}{G337.1}
\newcommand{\firdcthreethreesevenptwo}{G337.2}

\title{Infrared dark clouds on the far side of the Galaxy}
\author{
A. Giannetti \inst{\ref{mpi}}
\and F. Wyrowski \inst{\ref{mpi}}
\and S. Leurini \inst{\ref{mpi}}
\and J. Urquhart \inst{\ref{mpi}}
\and T. Csengeri \inst{\ref{mpi}}
\and K.~M. Menten \inst{\ref{mpi}}
\and L. Bronfman \inst{\ref{unichile}}
\and F.~F.~S. van der Tak \inst{\ref{sron}, \ref{kapteyn}}
}
\institute{
Max-Planck-Institut f\"ur Radioastronomie, auf dem H\"ugel 69, D-53121, Bonn, Germany \label{mpi}
\and Departamento de Astronomía, Universidad de Chile, Casilla 36-D, Santiago, Chile \label{unichile}
\and SRON Netherlands Institute for Space Research, Landleven 12, 9747 AD Groningen, The Netherlands \label{sron}
\and Kapteyn Astronomical Institute, University of Groningen, The Netherlands \label{kapteyn}
}

\abstract{
  Infrared dark clouds are the coldest and densest portions of giant molecular clouds.
  The most massive ones represent some of the most likely birthplaces for the next generation of massive stars in the Milky Way.
  Because a strong mid-IR background is needed to make them appear in absorption, they are usually assumed to be nearby.
}
{
  We use THz absorption spectroscopy to 
  solve the distance ambiguity associated with kinematic distances for the IR-dark clouds
  in the TOP100 ATLASGAL sample, a flux-limited selection of massive clumps in different evolutionary phases of star formation.
}
{
  The para-H$_2$O ground state transition at $1113.343\usk\giga\hertz$, observed with Herschel/HIFI, was used to investigate the occurrence of foreground absorption along the line of sight directly towards infrared-dark clouds. Additional consistency checks were performed using MALT90 and HiGAL archival data and targeted Mopra and APEX spectroscopic observations.
}
{
  We report the first discovery of five IRDCs in the TOP100 lying conclusively at the far kinematic distance, showing that the mere presence of low-contrast mid-IR absorption is not sufficient to unequivocally resolve the near/far ambiguity in favour of the former.
  All IRDCs are massive and actively forming high-mass stars; four of them also show infall signatures.
}
{
  We give a first estimate of the fraction of dark sources at the far distance ($\sim11\%$ in the TOP100) and describe their appearance and properties.
  The assumption that all dark clouds lie at the near distance may lead, in some cases, to underestimating masses, sizes, and luminosities, possibly causing clouds to be missed that will form very massive stars and clusters.
}

\keywords{ISM: clouds, ISM: lines and band, Stars: formation, Submillimeter: ISM}
\offprints{A. Giannetti (agiannetti@mpifr-bonn.mpg.de)}

\begin{document}  

\maketitle\ 

\section{Introduction}\label{sec:intro}

  Infrared dark clouds (IRDCs) were first discovered by the Infrared Space Observatory and the Midcourse Space Experiment as dark regions of absorption against diffuse mid-IR background emission \citep[e.g.,][]{Perault+96}. 
  They are ideal targets for studying the earliest stages of the process of star formation, because they host complexes of cold ($T\lesssim25\usk\kelvin$), dense ($n\gtrsim\pot{5}\cm^{-3}$), and massive ($M\gtrsim100\msun$) clumps. 
  The most massive and dense IRDCs are associated with high-mass star formation \citep[e.g.,][]{Pillai+06,Giannetti+13_aa556_16}, and they appear to contain a large fraction of the star-forming molecular gas in our Galaxy \citep{KauffmannPillai10_apjl723_7}.
  IRDCs, however, are not exceptional clouds, but rather they constitute the densest parts of much more extended giant molecular clouds \citep[GMCs; e.g.,][]{Teyssier+02_aa382_624,Schneider+15_aa758_29}, and some of them have also been discovered in the outer Galaxy \citep[e.g., ][]{Frieswijk+08_apjl685_L51}.
  
  For a cloud to be identified as an IRDC in mid-IR continuum images, the presence of background radiation is a necessary condition.
  It is commonly assumed \citep[e.g.,][]{Simon+06,Rathborne+06} that IRDCs should mainly lie at the near distance, where the background is strong enough 
  to make them appear clearly in absorption. On the other hand, it is possible that, in some cases, the background could be produced locally by, for example, nearby, more evolved star-forming regions. 
  In such cases some IRDCs could also lie at the far distance. 
  
  To derive the distribution and the basic properties of molecular clouds, such as their mass, density, size, and luminosity, distance determinations are fundamental. 
  Whereas \hi\ absorption measurements need a background source of continuum radiation \citep[e.g.,][]{Fish+03_apj587_701},
  molecular clouds are very bright at far-IR wavelengths, and even most IRDCs emit considerably in this regime, producing the background emission themselves. Molecular transitions with an excitation energy lower than is associated with the background continuum radiation will appear in absorption. 
  Absorption spectroscopy in the THz regime can therefore be used to break the distance ambiguity typical of kinematic distances in the inner Galaxy, even for cold and dark clouds.  
  \hi\ self-absorption is a reliable method of identifying nearby objects, when an \hii\ region is missing, but it is not as accurate for sources at the far kinematic distance \citep[][and references therein]{Wienen+15_aa579_91}.

  In this Letter we present observations of the H$_2$O$(1_{11}-0_{00})$ transition in five IRDCs, performed with HIFI \citep{deGraauw+10} on board the Herschel\footnote{Herschel is an ESA space observatory with science instruments provided by European-led Principal Investigator consortia and with important participation from NASA.}
  space observatory and complemented by Mopra, IRAM \nobrkdash{30}m and APEX data, to confirm that they lie at the far kinematic distance.

\section{Observations and sample of IRDCs}\label{sec:obs_and_sample}
  
  The APEX Telescope Large Area Survey of the Galaxy \citep[ATLASGAL, ][]{Schuller+09} is the first complete survey of the inner Galactic plane, and it was carried out with the Atacama Pathfinder Experiment \nobrkdash{12}m telescope (APEX) in the sub-mm continuum at $\lambda=870\mum$ ($\theta_\text{beam} = 19.2\asec$).
  The brightest sub-mm sources (TOP100 sample) were selected according to the criteria described in \citet{Giannetti+14_aa570_65}, so as to include different evolutionary stages, from quiescent clumps to \hii\ regions. The basic properties of all the sources in this sample are described in the same work and in K\"onig et al. (subm.). The IR-dark clumps are 45: their average flux density at $8\mum$ or $24\mum$ within the ATLASGAL beam is lower than the average flux at the same $\lambda$ in their vicinity. 
    \begin{figure}[t]
      \centering
      \includegraphics[width=0.8\columnwidth]{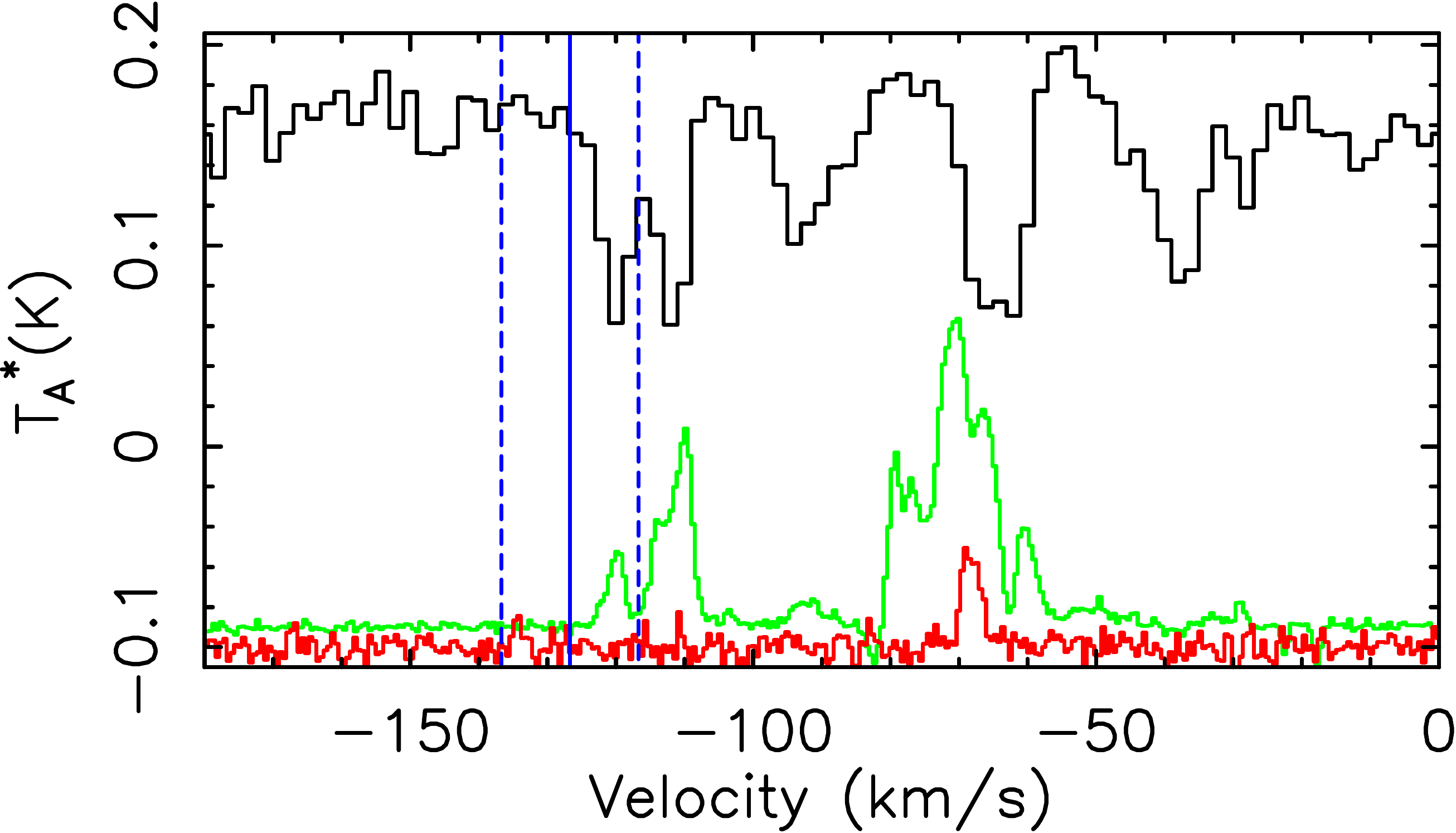}
      \caption{\emph{Black:} HIFI spectrum showing the para-H$_2$O ground-state absorption line for the line of sight towards \firdcthreethreesevenpone. \emph{Green:} FLASH CO$(3-2)$ spectrum ($1/50$x, shifted by $-0.09\kel$). \emph{Red:} C$^{17}$O$(3-2)$ spectrum ($1/16$x, shifted by $-0.11\kel$). The tangential velocity ($\pm10\kms$, dashed) is indicated in blue.}\label{fig:hifi_water} 
    \end{figure}

  Single-pointing observations for the sources in this sample were performed with Herschel/HIFI for lines from water \citep{Leurini+14_aa564_11} and for other molecules with the IRAM \nobrkdash{30}m telescopes at $90\usk\text{GHz}$ (Csengeri et al., 2015, subm.), APEX, and Mopra \nobrkdash{22}m (\citealt{Giannetti+14_aa570_65}, Giannetti et al., in prep., Wyrowski et al., in prep.). The molecular transitions used in this work, their frequencies, and the corresponding angular resolution are listed in Table~\ref{tab:obs_trans}.  
  
  Overall, $34$ ($\sim76\%$) dark clouds are detected in absorption. The water observations for the full TOP100 sample will be presented in a subsequent paper.
  The IRDCs studied here are G010.4446$-$0.0178 (hereafter \firdconezero\ for brevity), G018.7344$-$0.2261 (\firdconeeight) in the first quadrant, and G336.9574$-$0.2247 (\firdcthreethreesix), G337.1751$-$0.0324 (\firdcthreethreesevenpone), and G337.2580$-$0.1012 (\firdcthreethreesevenptwo) in the fourth quadrant. The last two clouds appear in the \citet{PerettoFuller09_aa505_405} catalogue, whereas \firdconezero\ and \firdcthreethreesix\ are not included there, despite having a similar contrast, because they lie in a region with a strong gradient in the background emission. On the other hand, \firdconeeight\ probably has a contrast that is too low to appear in the catalogue.
  The coordinates and the source velocities are listed in Table~\ref{tab:distances_coo}, and an overview of the properties of the sources is given in Table~\ref{tab:cloud_properties}.

  Four sources are covered by the MALT90 survey, which mapped more than 2000 objects in 16 molecular lines, including HCO$^+(1-0)$ and HNC$(1-0)$, with Mopra at an angular resolution of $\sim40\asec$. The observations and the data reduction process are described in \citet{Foster+11_apjs197_25,Foster+13_pasa30_38} and \citet{Jackson+13_pasa30_57}.   

\section{Results}

    The systemic LSR velocities of the sources in the TOP100 sample ($V_\text{cloud}$) are known from observations of the C$^{17}$O$(3-2)$ line \citep{Giannetti+14_aa570_65}. Assuming a rotation curve for the Galaxy, the near and far kinematic distances can be estimated. 
    If there are other molecular clouds between the observer and a massive cloud, which is bright at THz frequencies, they will absorb the continuum radiation at their respective velocity. In particular, when the massive cloud lies at the far distance, absorption features from foreground objects may have a systemic velocity between $|V_\text{cloud}|$ and the tangential velocity, depending on their positions.
    As an example, in the following we use \firdcthreethreesevenpone, and Figure~\ref{fig:hifi_water} shows its water spectrum. Clear absorption features are visible with velocities that are more negative than for the IRDCs, near $\sim-70\kms$, strongly suggesting that this cloud lies at the far distance.
    The spectra of the other objects are shown in Fig.~\ref{fig:hifi_water_online}. 
    The remaining 40 sources in the TOP100 do not show water absorption at velocities between that of the cloud and the tangential velocity, and thus are very likely to be at the near distance.

\begin{figure}[t]
      \centering
      \includegraphics[width=0.75\columnwidth]{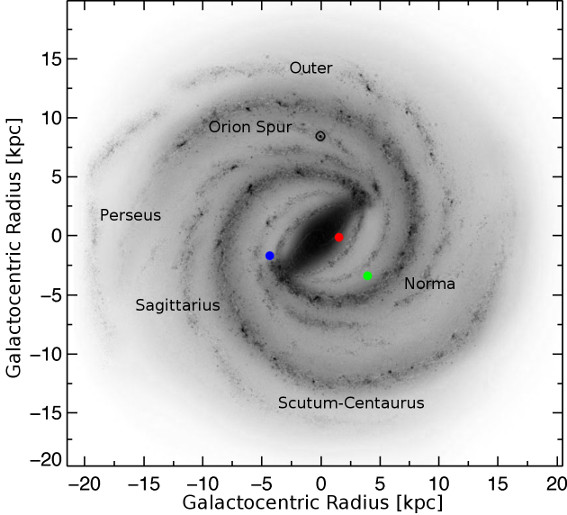}
      \caption{Position of the IRDCs in the Galaxy (Courtesy of NASA/JPL-Caltech): \firdcthreethreesix, \firdcthreethreesevenpone, and \firdcthreethreesevenptwo\ are indicated by a blue circle, \firdconezero\ and \firdconeeight\ by a red and a green one. The position of the Sun is also indicated.}\label{fig:position_galaxy}
    \end{figure}

    The rotation curve from \citet{Reid+14_apj783_130} (model B1)
    and from \citet{BrandBlitz93} give consistent results for all sources, with differences $\lesssim0.3\kpc$.
    \firdcthreethreesix, \firdcthreethreesevenpone, and \firdcthreethreesevenptwo\ should be located at $\sim11\kpc$ from the Sun, close to the far tip of the long Bar (Fig.~\ref{fig:position_galaxy}, blue circle)
    \firdconezero\ (Fig.~\ref{fig:position_galaxy}, red circle) could be associated to G010.4722$+$0.0277, because of its vicinity ($\sim3^\prime$) and of the similar $\vlsr$. \citet{Sanna+14_apj781_108} put this source in the connecting arm at $8.55^{+0.63}_{-0.55}\kpc$ from measurements of the methanol maser parallax, so we assume this distance. Finally, \firdconeeight\ (Fig.~\ref{fig:position_galaxy}, green circle) is $\sim12.5\kpc$ away from the Sun between the Sagittarius and the Perseus arms. 
    
\begin{figure}[t]
	\centering
	\includegraphics[width=0.75\columnwidth]{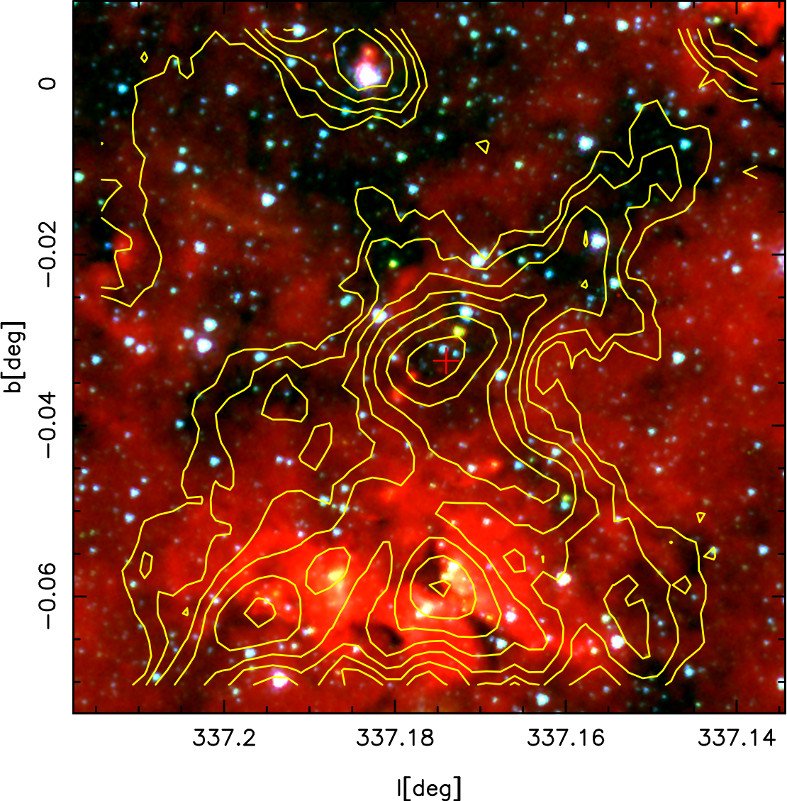}
	\caption{GLIMPSE false-colour image (red: $8\mum$; green: $4.5\mum$; blue: $3.6\mum$) of \firdcthreethreesevenpone. The yellow contours show the $870\mum$ emission, ranging from $0.281$ to $2.24\jybeam$ in equal log steps of $0.15$~dex. The red plus indicates the position of the molecular-line observations.}\label{fig:glimpse}
    \end{figure}

    In Figures~\ref{fig:glimpse} and \ref{fig:glimpse_online} the $8.0\mum$, $4.5\mum$, and $3.6\mum$ images from GLIMPSE are combined in a three-colour image, showing the mid-IR absorption associated with the $870\mum$ emission (contours). As mentioned in Sect.~\ref{sec:intro}, substantial background mid-IR emission must be present for the cloud to appear in absorption. All of the clouds considered here are associated with regions that have recently undergone intense star formation: the aforementioned G010.4722$+$0.0277 source is a massive proto-cluster candidate in the TOP100 sample \citep{Ginsburg+12_apj758_29,Urquhart+13_mnras431_1752}, with a luminosity $\sim5\times\pot{5}\lsun$. \firdconeeight\ lies in front of the Perseus arm, and several \hii\ regions are found in its vicinity, consistent with its distance \citep[e.g., G018.832$-$0.300][]{Anderson+12_apj754_62}. \firdcthreethreesix, \firdcthreethreesevenpone, and \firdcthreethreesevenptwo\ are associated with the far tip of the Bar and are discussed in more detail in Sect.~\ref{ssec:far_side_bar}.
    The presence of more evolved regions of star formation around the dark clouds supports the hypothesis that the background continuum emission at mid-IR wavelengths may be produced locally, for instance, by small dust grains heated by nearby \hii\ regions, before being absorbed by the intervening cold cloud. 

\section{Discussion}
  \subsection{Questioning the near distance}
    The presence of foreground absorption in the water spectra along the line of sight towards the IRDCs can be only explained if the cloud is located at the far distance. 
    In the following we show that alternative configurations are not consistent with the observations: (a) The dark clouds are not associated with the absorption at $V_\text{s}$ (the systemic LSR velocity of the brightest emission of high-density tracers  and C$^{17}$O). Another massive cloud lies behind it with the IRDC in the foreground, corresponding to one of the other absorption features.
    (b) The IRDCs have a velocity of $V_\text{s}$ and are at the near distance. The THz continuum would then be produced by another massive cloud in the background.

    Case (a) can be dismissed readily because the morphology of the emission at $V_\text{s}$ from MALT90 is consistent with the peak of the $870\mum$ emission and with the mid-IR absorption (see Fig.~\ref{fig:malt90_overlay}), whereas the other features (when detected) originate in different locations. 
    \firdconeeight\ was not observed in MALT90, so we cannot use this diagnostic.
    Additional evidence comes from the weak emission of high-density tracers, such as HCO$^+(1-0)$ or HNC$(1-0)$, for different velocity components even in the deep observations carried out with Mopra and APEX (Fig.~\ref{fig:deep_spectra_online}), and from the association of CH$_3$OH maser emission with $V_\text{s}$.

    Case (b) can also be ruled out because the (optically thin) C$^{17}$O$(3-2)$ emission in Figs.~\ref{fig:hifi_water} and \ref{fig:hifi_water_online} clearly shows that the dark clouds have by far the highest column densities among the clouds in these lines of sight, so they are the major contributors to the continuum emission. Indeed, the morphology of the continuum emission in the Herschel/HiGAL images at $250\mum$ still matches the IR absorption. 
    The minimum $N(\text{H}_2)$ needed to produce the continuum required to observe the water absorption features in the HIFI spectra is a significant fraction of the total one (Table~\ref{tab:cloud_properties}), and from C$^{17}$O, we can see that only the IRDCs have enough column density to produce such strong continuum.
    Beam dilution could play an important role in suppressing the emission from high-density tracers in a possible massive background source. To exclude this possibility we used the emission from CO which most likely fills the APEX beam and is proportional to the mass of the cloud \citep[e.g.,][]{Garcia+14_apjs212_2}: indeed, the CO$(3-2)$ emission (Fig.~\ref{fig:hifi_water}) is weaker for the absorption features at different velocities from that of the IRDC, indicating that it is the most massive source along the line of sight.
    
    Despite their very limited number, these objects share a couple of traits: the contrast of the absorption against the background is low (cf. Sect.~\ref{sec:obs_and_sample}), as would be expected, for example, because of strong foreground, emission and they have clearly defined, compact, and relatively isolated peaks in the sub-mm emission. Therefore, at least when dealing with sources with these characteristics, care must be taken in assigning distances.
    The five IRDCs considered in this work also suggest that the contrast in mid-IR images is not directly connected to column density, a result discussed in more detail in Urquhart et al. (in prep.). These authors cross-correlate emission in the ATLASGAL survey with the catalogue of \citet{PerettoFuller09_aa505_405}, finding that only a small fraction of the IRDCs candidates have a sub-mm counterpart.

  \subsection{Gravitational stability and star formation}
    
    K\"onig et al. (subm.), construct the SEDs for the sources in the TOP100 sample deriving dust temperatures and bolometric luminosities. With this dust temperature, we recomputed the masses (Table~\ref{tab:cloud_properties}) from the sub-mm flux given in \citet{Csengeri+14_aa565_75}. 
    Using the sizes and linewidths reported in \citet{Giannetti+14_aa570_65}, we also recomputed the virial parameter $\alpha = \mvir/M$.
    The embedded clumps appear to be gravitationally bound with values of $\alpha \lesssim 1$ (Table~\ref{tab:cloud_properties}).

    We find that four sources have spectra that are typical of contraction in the high-density tracers and in CO$(3-2)$ (Figures~\ref{fig:hifi_water}, \ref{fig:hifi_water_online}, and \ref{fig:deep_spectra_online}): an asymmetric line profile in optically thick lines with the blue wing stronger than the red one \citep[see ][]{Zhou+93_apj404_232,Gregersen+97_apj484_256}. 
    Another piece of evidence comes from the water line, which is redshifted with respect to the systemic velocity \citep[Figs.~\ref{fig:hifi_water} and~\ref{fig:hifi_water_online}; cf. discussion in][]{Wyrowski+12_aa542_15,Shipman+14_aa570_51}. The spectra are crowded, the spectral resolution is low, and the lines are broad and possibly saturated. It is thus difficult to have an accurate estimate of the velocity shift; considering the uncertainties, Gaussian fits to the absorption lead to shifts comparable to the infall velocity derived from HCO$^+$ (Table~\ref{tab:cloud_properties}).
    \citet{Kauffmann+13_apj779_185} show that it is possible for a collapsing cloud not to have $\alpha\ll1$ if energy is conserved and the cloud has undergone significant contraction.

    Assuming that the clouds are infalling, we can derive an upper limit for the mass infall rate $\dot{M}_{in}$ using Eq.~5 in \citet{LopezSepulcre+10_aa517_66}, where the infall velocity $V_{in}$ is estimated according to Eq.~9 in \citet{Myers+96}: $\dot{M}_{in}$ (listed in Table~\ref{tab:cloud_properties}) is found to be in the range $2-18\times\pot{-3}\msun\yr^{-1}$.
    The infall rates that we find are similar to those of other high-mass star-forming regions \citep[e.g., ][]{LopezSepulcre+10_aa517_66}.
    The infall velocities and rates for the IRDCs studied in this Letter are also comparable to the one given by \citet{Peretto+13_aa555_112}. These authors investigate the infall in a massive star-forming IRDC with a total mass comparable to that contained in the compact sub-mm peak of the sources investigated here, and spatially resolve the collapse, showing that the cloud is undergoing global contraction.

    The presence of class II methanol masers \citep[][]{Urquhart+13_mnras431_1752} is proof that high-mass stars are being formed in these IRDCs, in agreement with empirical thresholds for massive star formation \citep{KauffmannPillai10_apjl723_7,Urquhart+13_mnras431_1752}. 
    \firdconeeight\ and \firdcthreethreesevenpone\ are found to be near the massive proto-cluster locus, using the same assumptions as in \citet{Urquhart+13_mnras431_1752}.
    Considering a star formation efficiency of $\sim30\%$ \citep[e.g., ][]{Urquhart+13_mnras431_1752}, we obtain an order-of-magnitude estimate for the stellar mass of the cluster that is being formed between $500\msun$ and $2200\msun$. This implies $M_*\approx13-40\msun$ for the most massive member, which was obtained using a \citet{Kroupa02_sci295_82} initial mass function with $\alpha_3 = 2.7$ for masses above $1\msun$.

  \subsection{Star formation activity on the far side of the Bar} \label{ssec:far_side_bar}
  
    There is increasing evidence that the region around the far side of the Bar has recently undergone a strong event of star formation. Several \hii\ regions have been found in this region, for example G337.0$+$0.0 \citep{Russeil03} and G337.12$-$0.18 \citep{Wienen+15_aa579_91}, at $\sim11\kpc$ from the Sun from \hi\ absorption measurements. Moreover, \citet{Russeil03} finds a group of giant \hii\ regions in this direction. 
    Here \citet{Garcia+14_apjs212_2} find two of the most massive (several $\pot{6}\msun$) and active GMCs in their survey of the IV quadrant, G336.875$+$0.125 and G337.750$+$0.000. The former could be enveloping \firdcthreethreesix, \firdcthreethreesevenpone, and \firdcthreethreesevenptwo, providing an enormous gas reservoir for star formation; it already hosts IRAS point sources with far-IR colours of UC\hii\ regions \citep[][]{Bronfman+96_aa115_81}. 
    Young massive clusters have been discovered on the far end of the Bar \citep[e.g., ][]{Davies+12}. 
    On the basis of these findings, \citet{Davies+12} suggests that the far side of the Bar may be as active as its near analogue, which is known to have recently undergone a starburst phase \citep[e.g.,][]{Garzon+97}. 
    The IRDCs studied here provide evidence that new massive clusters are also being born on the far side of the Bar.

\section{Summary and conclusions}

  In this Letter we have reported the first discovery of five IRDCs lying conclusively at the far kinematic distance, we gave a first estimate of the fraction of such objects and describe their appearance and properties.
  The association of these sources with more evolved regions of star formation supports the idea that the background in these cases could be produced locally. 
  In the TOP100 sample $\sim11\%$ of the dark sources are found to lie at the far kinematic distance, although one has to keep in mind that the sources are only classified on the basis of the brightest clump at sub-mm wavelengths, and some of the other sources may belong to an IR-dark complex.
  Therefore, assuming that all of the IRDCs are at the near distance may, for a small fraction of objects, lead to significantly underestimate the mass, size, and luminosity, possibly causing clouds to be missed that will form very massive stars and clusters.
  The IRDCs discussed in this Letter are found to have low contrast and compact, relatively isolated peaks in the sub-mm continuum, as can be expected if they lie at the far distance. Sources of this kind must therefore be treated with particular care, and more evidence is needed than the mere presence of mid-IR absorption to resolve the ambiguity in the kinematic distance.
  While now, after the end of the Herschel mission, the observation of the water ground-state absorption line is no longer possible, other molecules may be used instead. The ubiquitous OH$^+$ molecule \citep{Wyrowski+10_aa518_26} or ALMA observations towards compact continuum sources of transitions such as CO$(3-2)$ and HCO$^{+}(1-0)$, can be used to search for absorption features from mostly atomic or molecular clouds, respectively.

  We find convincing evidence that four of the five IRDCs are contracting, and given that these objects are far away, the probed linear scale is similar to the scale of the maps in \citet{Peretto+13_aa555_112} and \citet{Schneider+15_aa758_29}, possibly suggesting that the IRDCs in our sample are also globally infalling. All of them are actively forming massive stars, as revealed by their association with class II methanol maser emission \citep{Urquhart+13_mnras431_1752}. 
  The estimated mass infall rates are typical of high-mass star-forming regions, in the range $2-18\times\pot{-3}\msun\yr^{-1}$. 
  Using the same assumptions as in \citet{Urquhart+13_mnras431_1752} \firdconeeight\ and \firdcthreethreesevenpone\ fall near the massive proto-cluster candidate locus.
  
  Three of the five clouds (\firdcthreethreesix, \firdcthreethreesevenpone, and \firdcthreethreesevenptwo) are found to lie close to the far side of the Bar, where a few young clusters were discovered \citep[e.g.,][]{Davies+12}, showing that the process of star formation is vigorous in this region of the Galaxy. These clouds are not very different from the molecular ridges in W43 on the near of the Bar, similarly containing several thousand solar masses of material within a few parsecs \citep[e.g.,][]{NguyenLuong+13_apj775_88}, but possibly in an earlier evolutionary stage.

\begin{acknowledgements}
  We thank the referee for the comments that helped clarify and improve this paper.
  LB acknowledges support from CONICYT Basal Project PFB-06.
  HIFI has been designed and built by a consortium of institutes and university departments from across Europe, Canada, and the United States under the leadership of SRON Netherlands Institute for Space Research, Groningen, The Netherlands, and with major contributions from Germany, France, and the US. Consortium members are: Canada: CSA, U.Waterloo; France: CESR, LAB, LERMA, IRAM; Germany: KOSMA, MPIfR, MPS; Ireland, NUI Maynooth; Italy: ASI, IFSI-INAF, Osservatorio Astrofisico di Arcetri-INAF; Netherlands: SRON, TUD; Poland: CAMK, CBK; Spain: Observatorio Astronómico Nacional (IGN), Centro de Astrobiología (CSIC-INTA). Sweden: Chalmers University of Technology - MC2, RSS \& GARD; Onsala Space Observatory; Swedish National Space Board, Stockholm University - Stockholm Observatory; Switzerland: ETH Zurich, FHNW; USA: Caltech, JPL, NHSC.
\end{acknowledgements}

\bibliographystyle{bibtex/aa}
\bibliography{bibtex/biblio.bib}

\Online

\begin{appendix}
  \section{Online material}
    \begin{table*}
      \caption{Observed molecular transitions.}\label{tab:obs_trans}
      \centering
      \begin{tabular}{lrccc}
	\toprule
	\toprule
	Transition                               & Frequency  & Beam   & Telescope  & Instrument \\
	                                         & GHz        & arcsec &            &            \\
	\midrule                                                                                 
	H$_2$O$(1_{11}-0_{00})$\tablefootmark{a} & $1113.343$ & 19     & Herschel   & HIFI       \\
	HCO$^{+}(1-0)$                           & $89.189$   & 24-38  & 30m/Mopra  & EMIR/MOPS  \\
	H$^{13}$CO$^{+}(1-0)$                    & $86.754$   & 25-39  & 30m/Mopra  & EMIR/MOPS  \\
	HNC$(1-0)$                               & $90.664$   & 24-38  & 30m/Mopra  & EMIR/MOPS  \\
	HN$^{13}$C$(1-0)$                        & $87.091$   & 25-39  & 30m/Mopra  & EMIR/MOPS  \\
	CO$(3-2)$                                & $345.796$  & 18     & APEX       & FLASH      \\
	\bottomrule
      \end{tabular}
      \tablefoot{
      \tablefoottext{a}{The Observation IDs for the five sources are 1342268161, 1342266721, 1342263302, 1342263301, 1342263300.}
      }
    \end{table*} 

    \begin{table*}
      \caption{Coordinates and basic aspects of the five IRDCs.}\label{tab:distances_coo}
      \centering
      \begin{tabular}{l*{6}c}
	\toprule
	\toprule
	  Source                  & RA(J2000)     & DEC(J2000)    & $D$                     & $\reff$             & $L$                & $\vlsr$   \\      
	                          & HH:MM:SS      & DD:MM:SS      & $\kpctab$               & arcsec              & $\pot{3}\lsuntab$  & $\kmstab$ \\      
	\midrule                                                                                                                      
	\firdconezero             & $18:08:44.73$ & $-19:54:32.7$ & $8.55$\tablefootmark{a} & $23$                & $11.5$             & $74.8$    \\[3pt] %
	\firdconeeight            & $18:25:56.01$ & $-12:42:49.6$ & $12.5$                  & $28$                & $72.3$             & $40.8$    \\[3pt] %
	\firdcthreethreesix       & $16:36:17.03$ & $-47:40:49.8$ & $11.0$                  & $19$                & $3.5$              & $-71.3$   \\[3pt] %
	\firdcthreethreesevenpone & $16:36:18.43$ & $-47:23:25.0$ & $11.0$                  & $33$                & $61.0$             & $-68.2$   \\[3pt] %
	\firdcthreethreesevenptwo & $16:36:56.42$ & $-47:22:27.1$ & $11.0$                  & $25$                & $30.3$             & $-68.3$   \\[3pt] %
	\bottomrule
      \end{tabular}
      \tablefoot{Columns: right ascension, declination, distance, effective radius $\reff$ \citep[see ][]{Rosolowsky+10} for the compact emission identified in \citet{Csengeri+14_aa565_75}, luminosity from K\"onig et al. (subm.), and systemic LSR velocity of the source. 
      \tablefoottext{a}{Instead of the far kinematic distance, we adopt the distance measured with maser parallax for G010.4722$+$0.0277 (see text).}}
    \end{table*}

    \begin{table*}
      \caption{Overview of the physical properties of the IRDCs.}\label{tab:cloud_properties}
      \centering
      \begin{tabular}{l*{10}c}
	\toprule
	\toprule
	  Source                  & $\td$          & $M(\text{GC})$        & $N(\text{H}_2)$       & $N(\text{H}_2,min)$   & $n(\text{H}_2)$       & $\mvir$               & $\alpha$              & $V_{in}$  & $\Delta V_{\rm H_2O}$ & $\dot{M}_{in}$              \\
	                          & K              & $\pot{3}\msuntab$     & $\pot{22}\cm^{-2}$    & $\pot{22}\cm^{-2}$    & $\pot{3}\cm^{-3}$     & $\pot{3}\msuntab$     &                       & $\kmstab$ & $\kmstab$             & $\pot{-3}\msuntab \yr^{-1}$ \\
	\midrule                                                                                                                                                                                                                                        
	\firdconezero             & $20.8\pm{0.4}$ & $1.5\unc{-0.3}{+0.3}$ & $5.1\unc{-1.0}{+0.9}$ & $2.4$                 & $8.7\unc{-1.8}{+1.5}$ & $1.5\unc{-0.3}{+0.3}$ & $1.0\unc{-0.3}{+0.2}$ & $0.4$     & $1.9\pm1.0$           & $2$                         \\[3pt]
	\firdconeeight            & $22.0\pm{1.0}$ & $6.6\unc{-2.6}{+2.0}$ & $8.2\unc{-1.7}{+1.5}$ & $4.2$                 & $8.0\unc{-2.2}{+1.7}$ & $6.6\unc{-2.1}{+1.5}$ & $1.1\unc{-0.4}{+0.3}$ & $1.5$     & $3.5\pm0.9$           & $18$                        \\[3pt]
	\firdcthreethreesix       & $15.8\pm{1.0}$ & $2.2\unc{-1.0}{+0.6}$ & $5.8\unc{-1.4}{+1.1}$ & $3.6$                 & $9.5\unc{-3.0}{+2.0}$ & $1.8\unc{-0.9}{+0.5}$ & $0.9\unc{-0.5}{+0.3}$ & \dots     & \dots                 & \dots                       \\[3pt]
	\firdcthreethreesevenpone & $22.4\pm{0.9}$ & $4.0\unc{-1.8}{+1.1}$ & $5.2\unc{-1.0}{+1.0}$ & $3.2$                 & $4.9\unc{-1.4}{+1.1}$ & $1.6\unc{-0.6}{+0.4}$ & $0.4\unc{-0.2}{+0.1}$ & $2.4$     & $2.2\pm0.8$           & $17$                        \\[3pt]
	\firdcthreethreesevenptwo & $21.9\pm{1.1}$ & $2.8\unc{-1.2}{+0.8}$ & $5.3\unc{-1.1}{+1.0}$ & $3.8$                 & $6.6\unc{-2.0}{+1.4}$ & $2.2\unc{-0.9}{+0.7}$ & $0.9\unc{-0.4}{+0.3}$ & $0.5$     & $2.2\pm2.0$           & $3$                         \\[3pt]
	\bottomrule
      \end{tabular}
      \tablefoot{Cols.: dust temperature calculated by K\"onig et al., (subm.), mass calculated with the flux from \citet{Csengeri+14_aa565_75}, beam-averaged peak column density,  minimum H$_2$ column density needed to observe the water absorption features, derived from the fraction of continuum absorbed by water for the dark clouds (similar numbers also apply to the strongest absorption feature of the foreground clouds in each spectrum, based on their intensity), beam-averaged peak volume density, virial mass calculated from the C$^{17}$O$(3-2)$ linewidths listed in \citet{Giannetti+14_aa570_65}, virial parameter $\alpha = \mvir/M$, infall velocity from HCO$^+(1-0)$, redshift of the water lines with respect to the systemic velocity, and infall rates, calculated using $\rho = M(\text{GC}) / (4/3 \pi \reff^3)$.}
    \end{table*}
    
    \begin{figure*}[tbp]
      \includegraphics[width=\columnwidth]{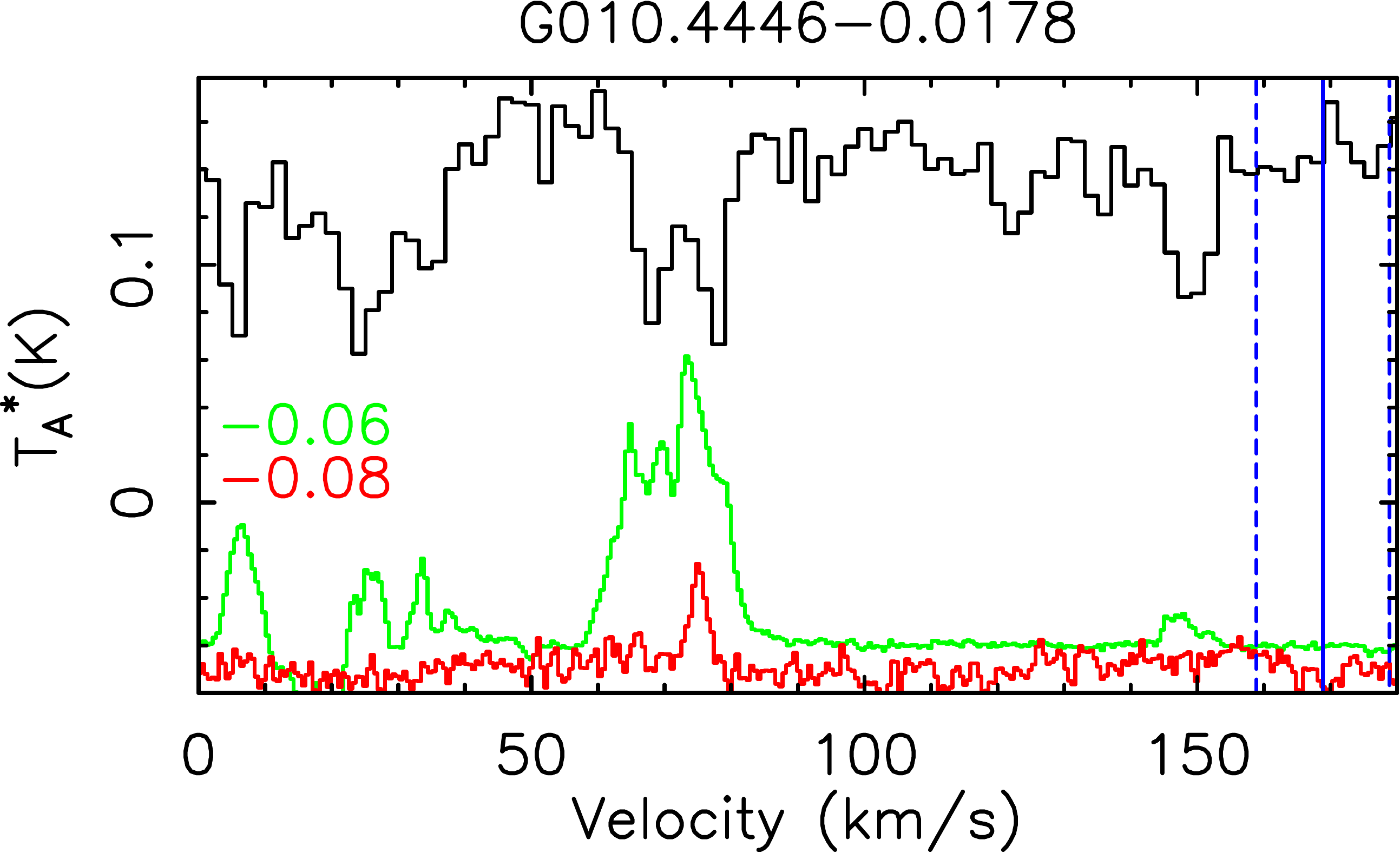}
      \includegraphics[width=\columnwidth]{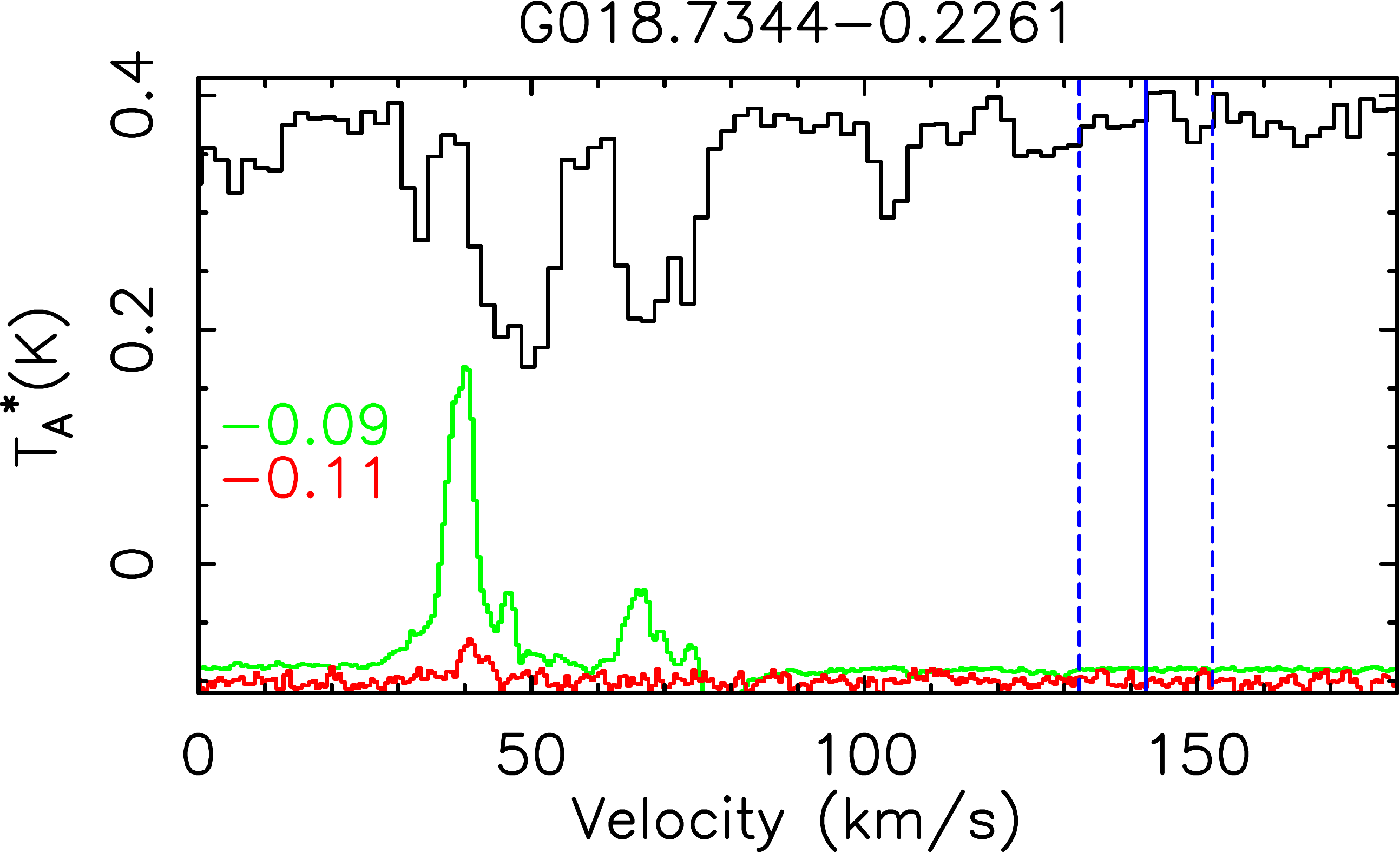}\hfill 
      
      \vspace*{0.5cm}
      
      \includegraphics[width=\columnwidth]{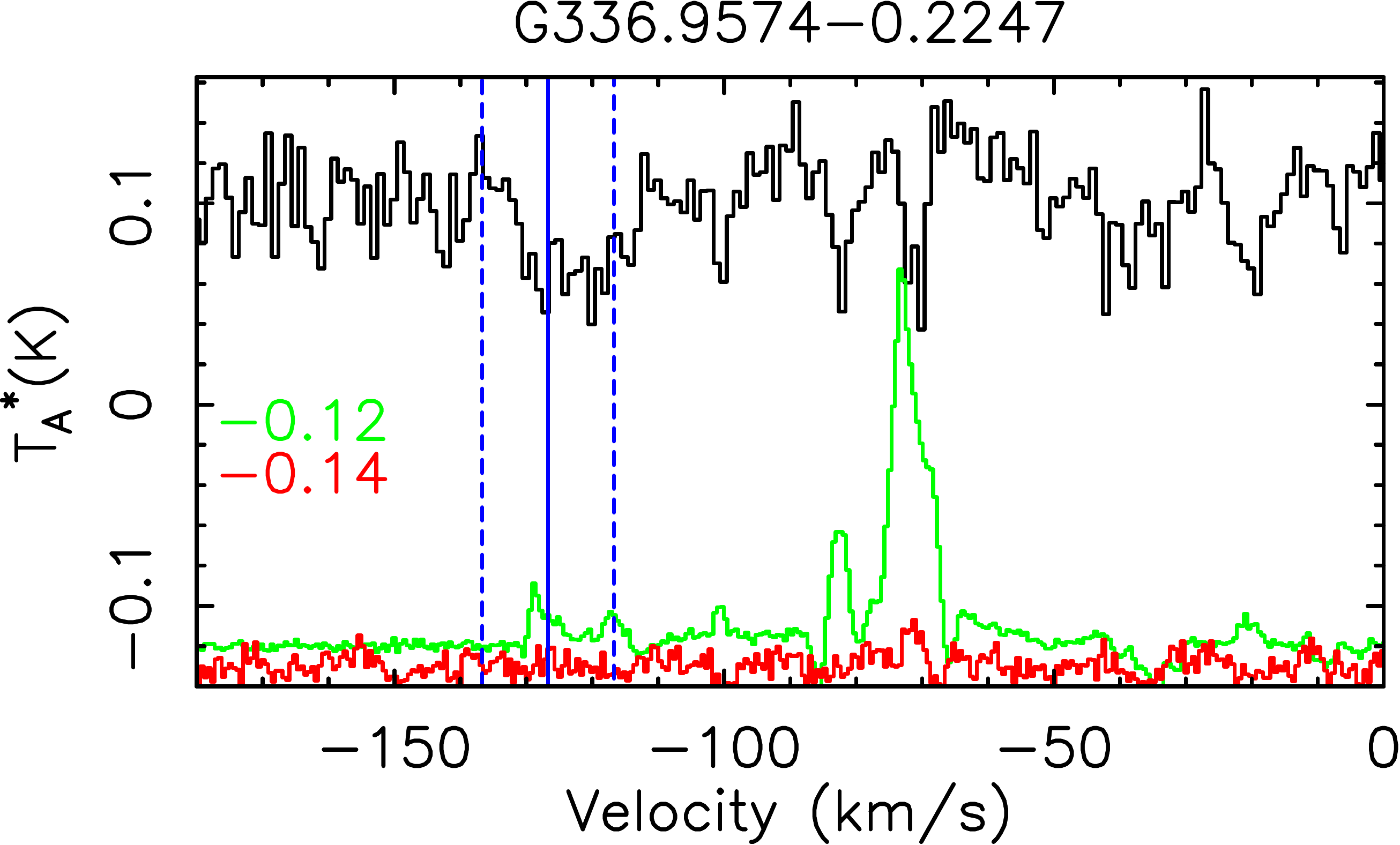}
      \includegraphics[width=\columnwidth]{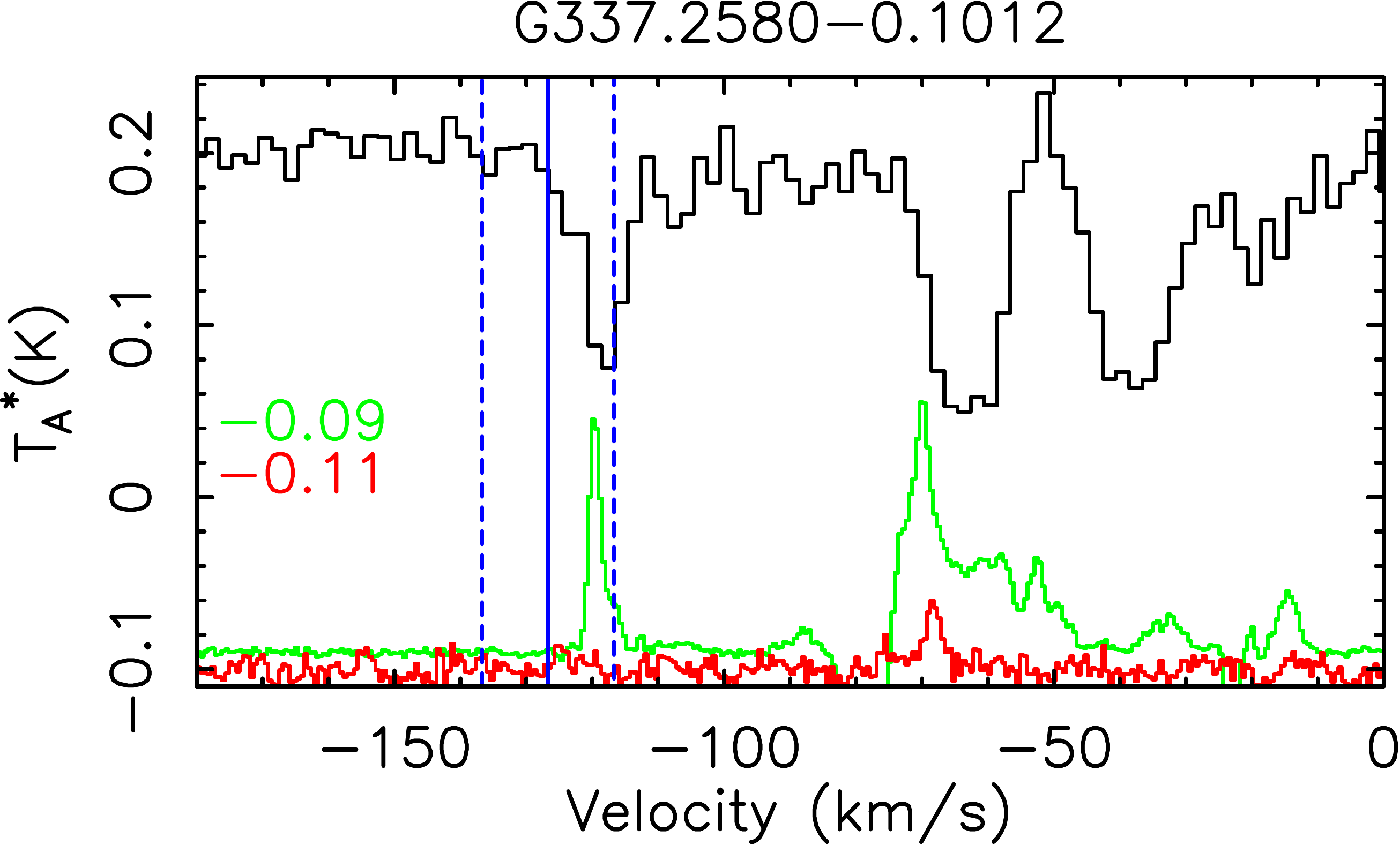}\hfill
      \caption{\emph{Black:} HIFI spectra of the para-H$_2$O ground-state absorption line at $1113.343\usk\giga\hertz$ for the line of sight towards \firdconezero\ (top left), \firdconeeight\ (top right), \firdcthreethreesix\ (bottom left), and \firdcthreethreesevenptwo\ (bottom right). \emph{Green:} FLASH CO$(3-2)$ spectra, rescaled dividing by 50 and shifted by the amount indicated in green on the left. \emph{Red:} C$^{17}$O$(3-2)$ spectrum, rescaled dividing by 16 and shifted by the amount indicated in red on the left. The tangential velocity ($\pm10\kms$, dashed) is indicated in blue.}\label{fig:hifi_water_online}
    \end{figure*}
    \begin{figure*}[t]
      \centering
      \includegraphics[width=0.9\columnwidth]{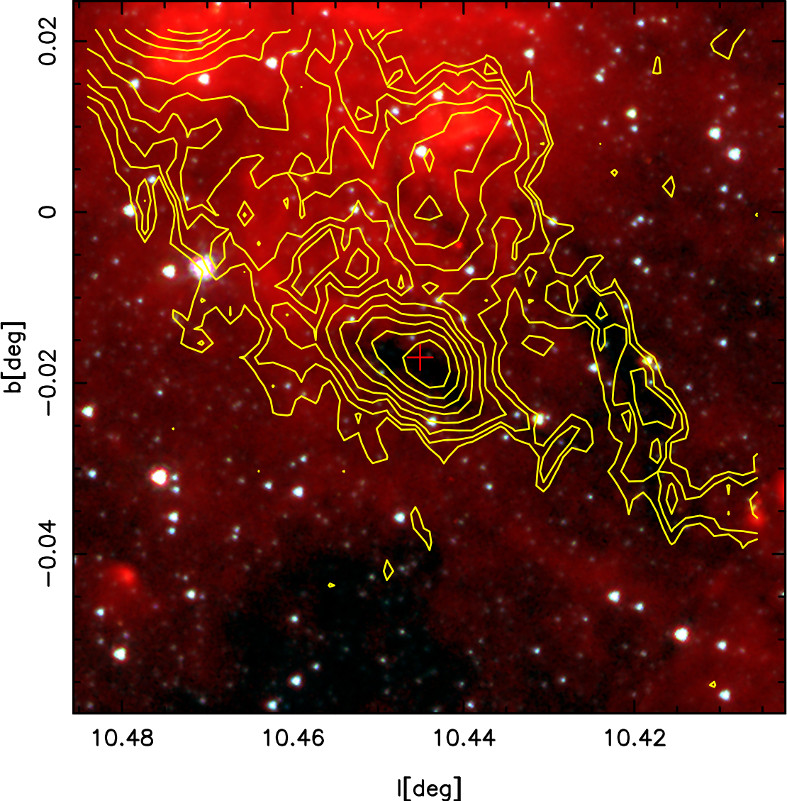}
      \hspace{0.07\columnwidth}
      \includegraphics[width=0.9\columnwidth]{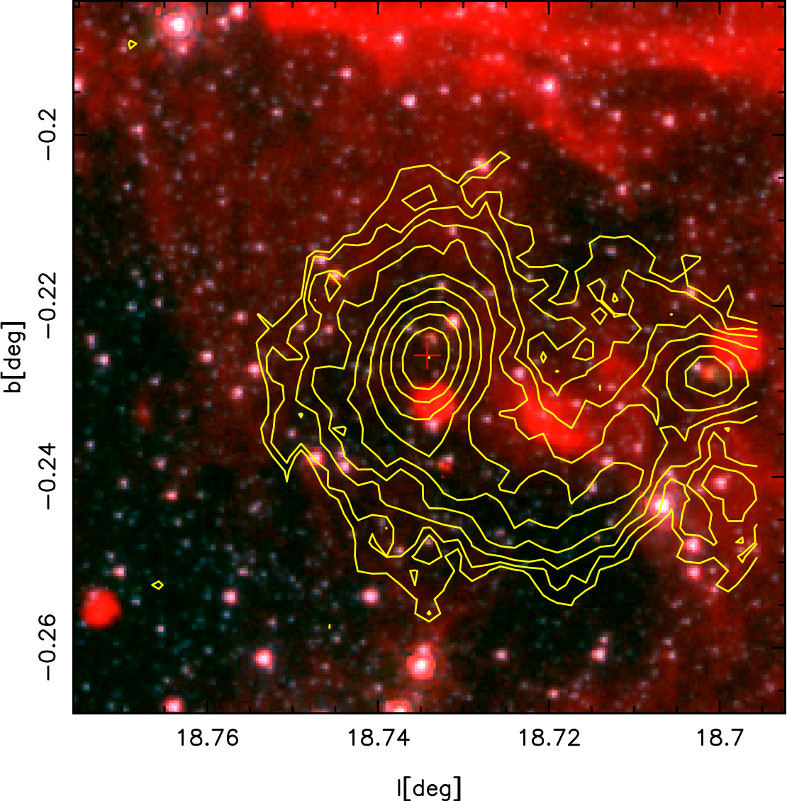} \\
      \vspace{1cm}
      \includegraphics[width=0.9\columnwidth]{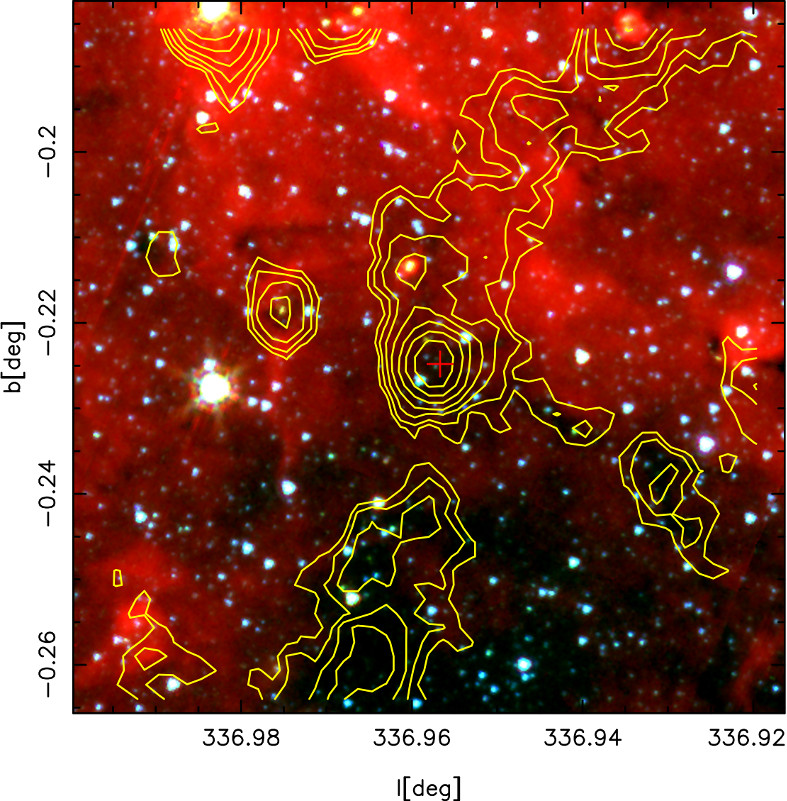}
      \hspace{0.07\columnwidth}
      \includegraphics[width=0.9\columnwidth]{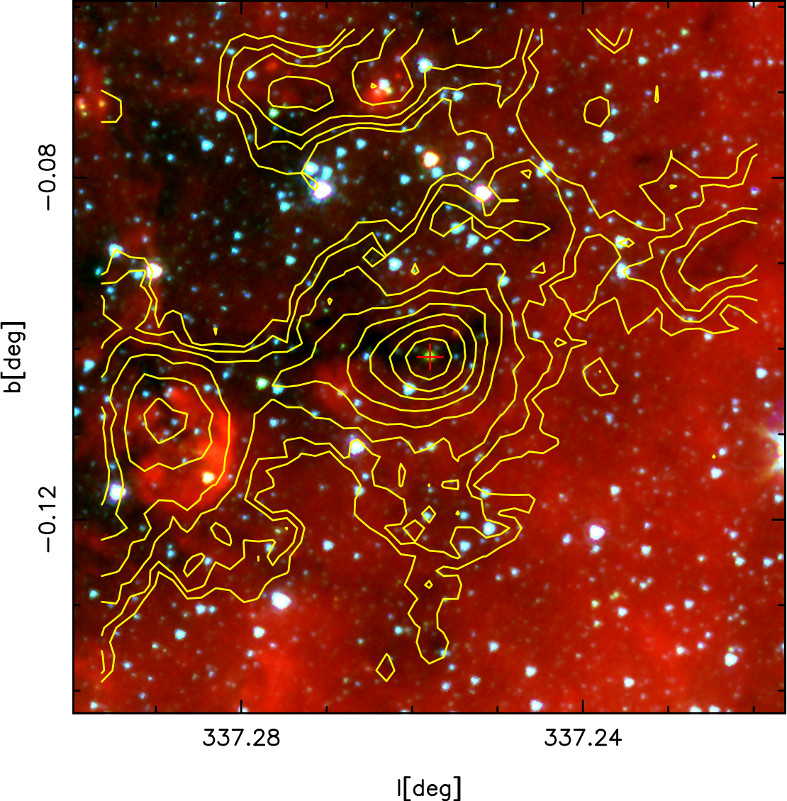}
      \caption{GLIMPSE false-colour images (red: $8\mum$, green: $4.5\mum$, blue: $3.6\mum$) of \firdconezero\ (a), \firdconeeight\ (b), \firdcthreethreesix\ (c), \firdcthreethreesevenptwo\ (d). The yellow contours show the $870\mum$ emission from the ATLASGAL survey, ranging from $0.14$ to $1.59\jybeam$ in equal log steps of $0.15$~dex in (a); from $0.14$ to $3.98\jybeam$ in equal log steps of $0.16$~dex in (b); from $0.14$ to $1.12\jybeam$ in equal log steps of $0.15$~dex in (c); and from $0.15$ to $1.91\jybeam$ in equal log steps of $0.16$~dex in (d). The red plus indicates the position of the molecular-line observations.}\label{fig:glimpse_online}
    \end{figure*}
    
    \begin{figure*}[tbp]
    \centering
      \includegraphics[width=0.9\columnwidth]{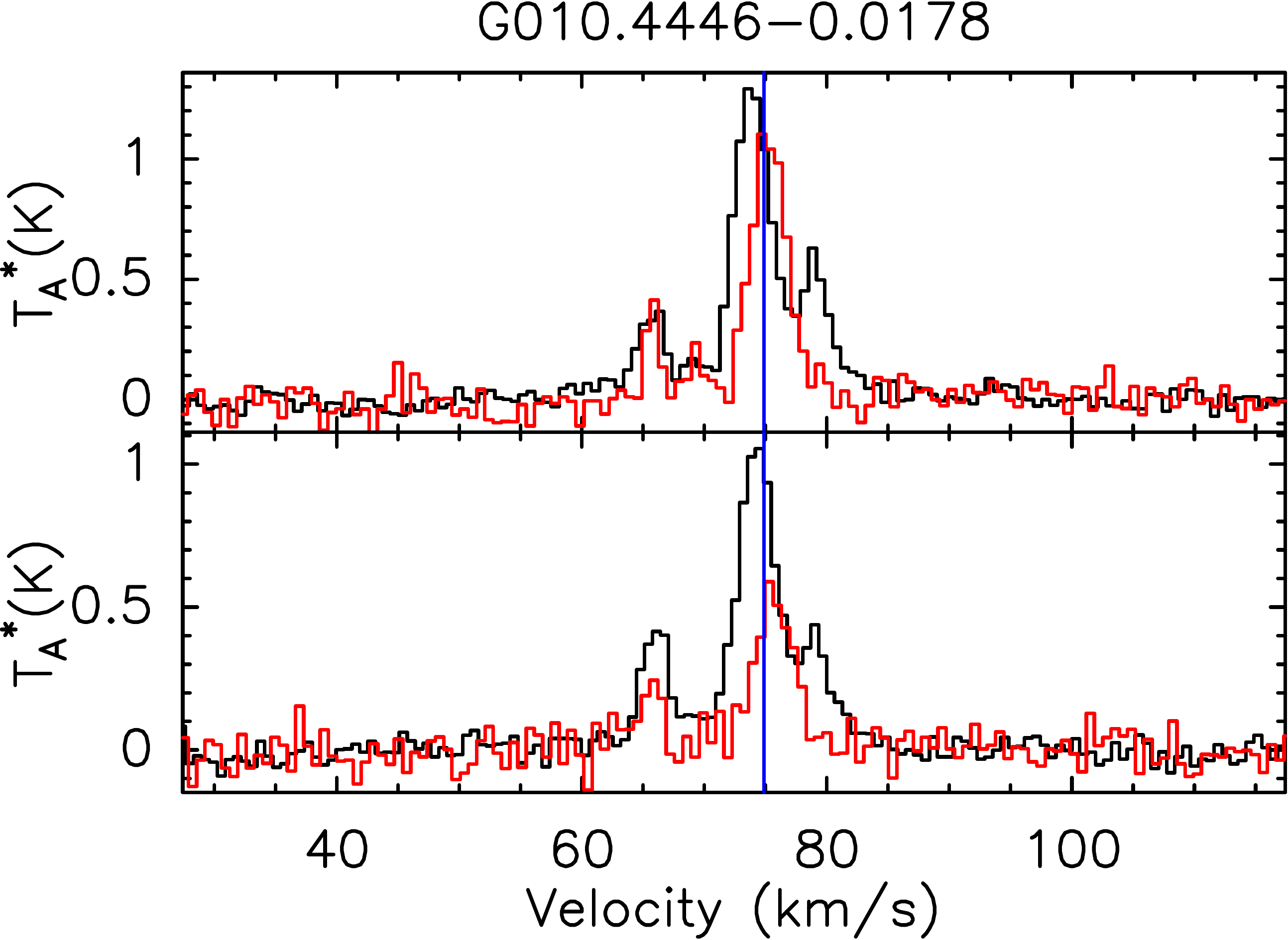}
      \hspace{0.07\columnwidth}
      \includegraphics[width=0.9\columnwidth]{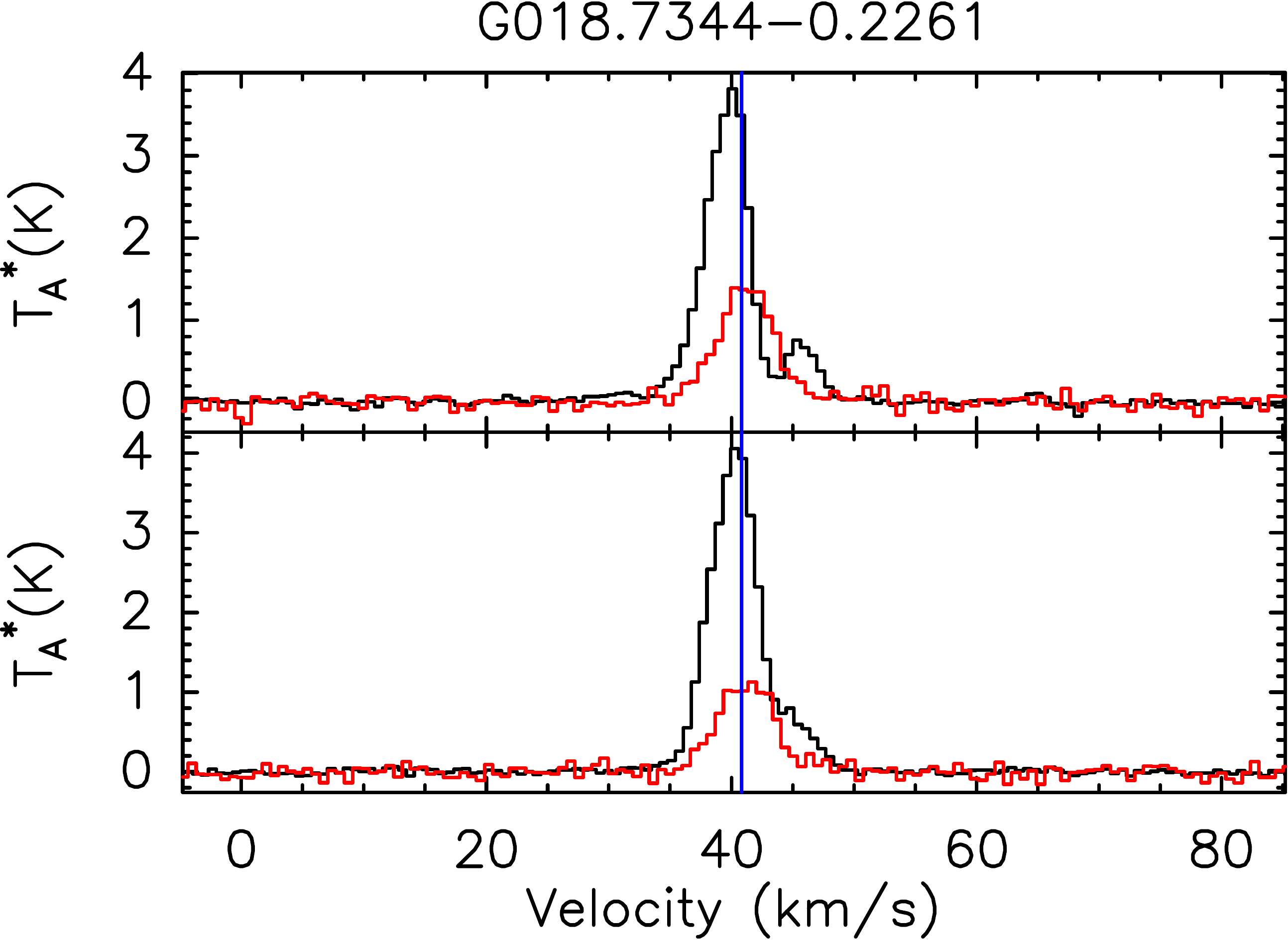} \\
      \vspace{1cm}
      \includegraphics[width=0.9\columnwidth]{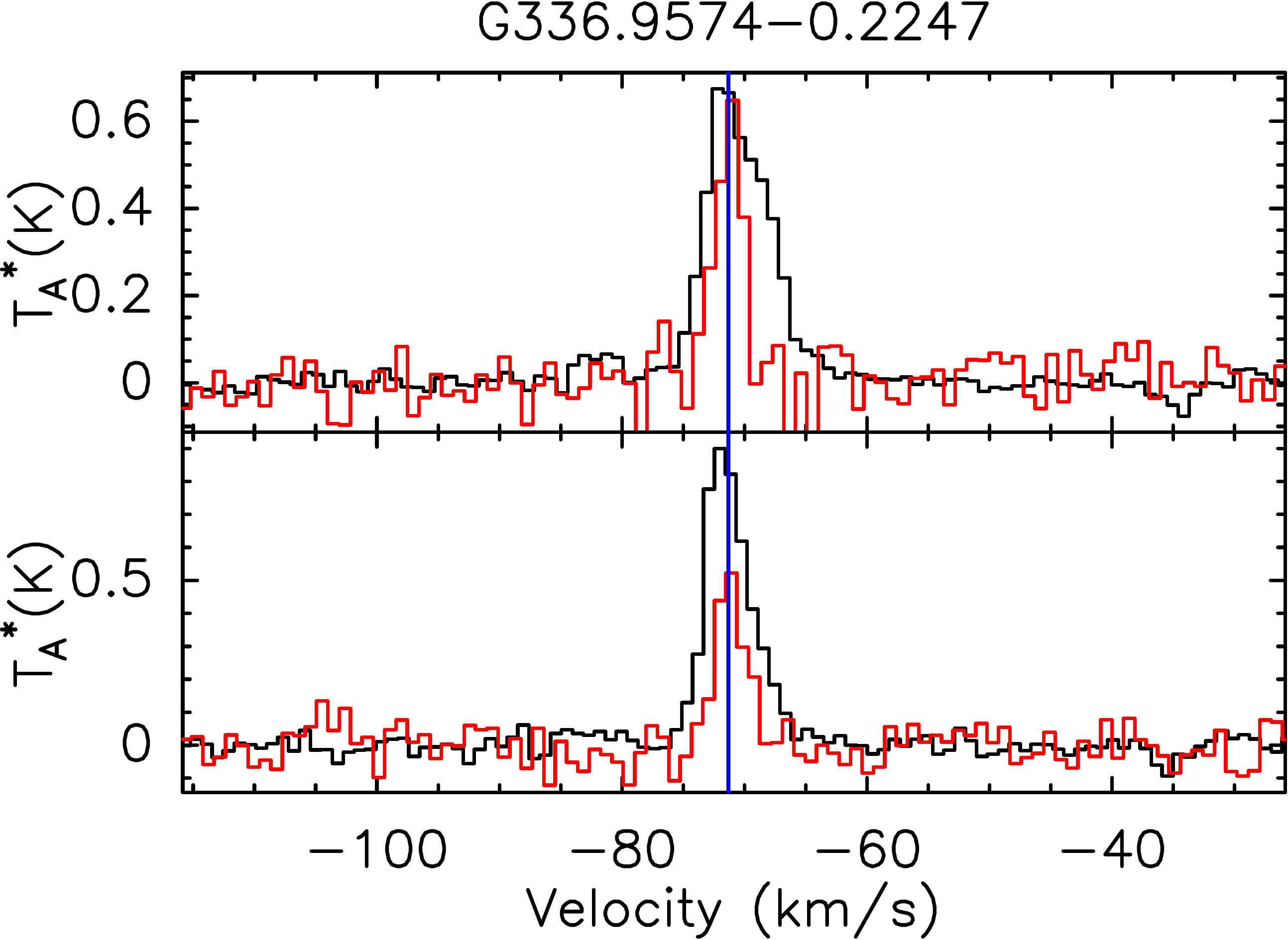}
      \hspace{0.07\columnwidth}
      \includegraphics[width=0.9\columnwidth]{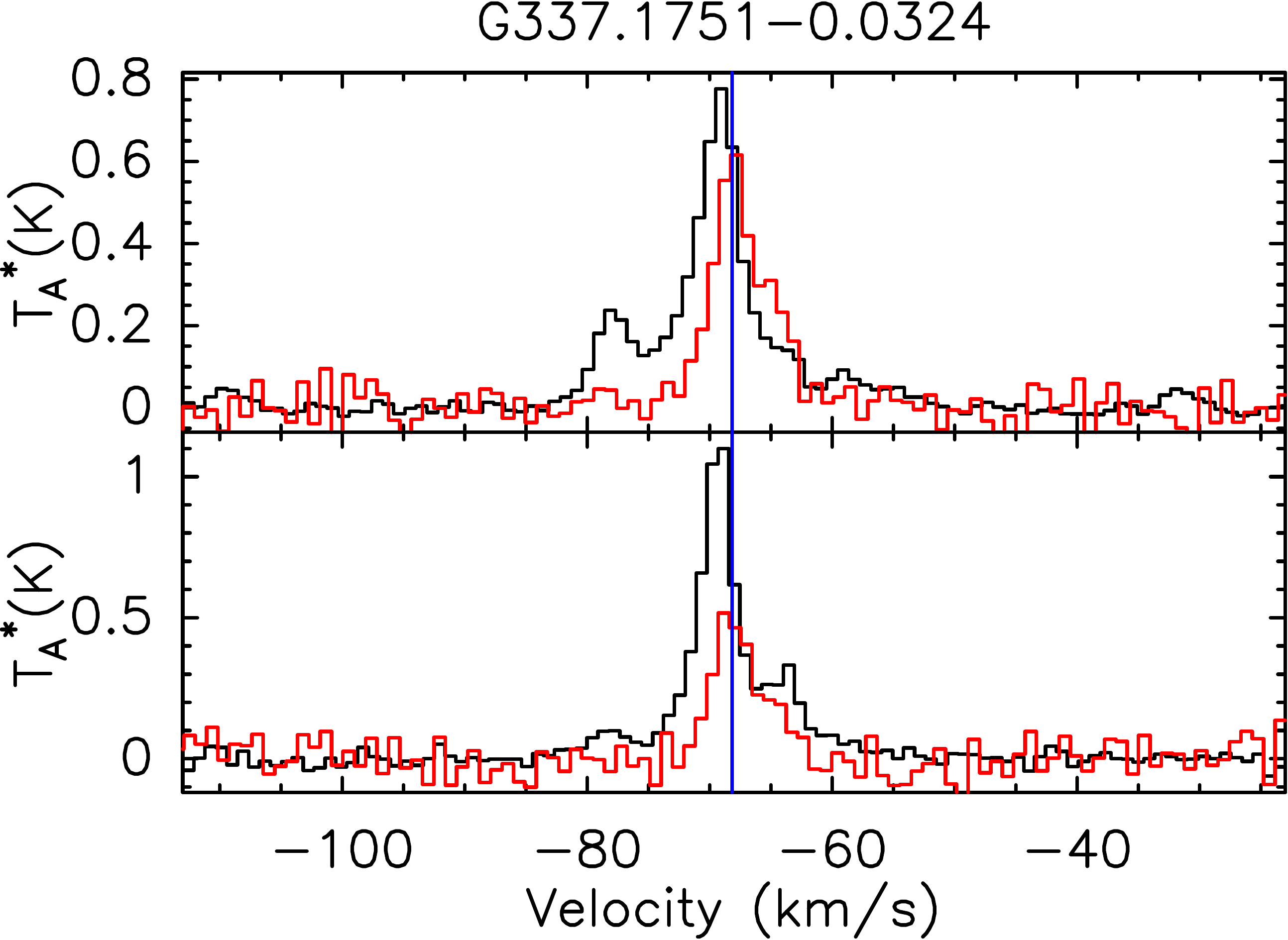} \\
      \vspace{1cm}
      \includegraphics[width=0.9\columnwidth]{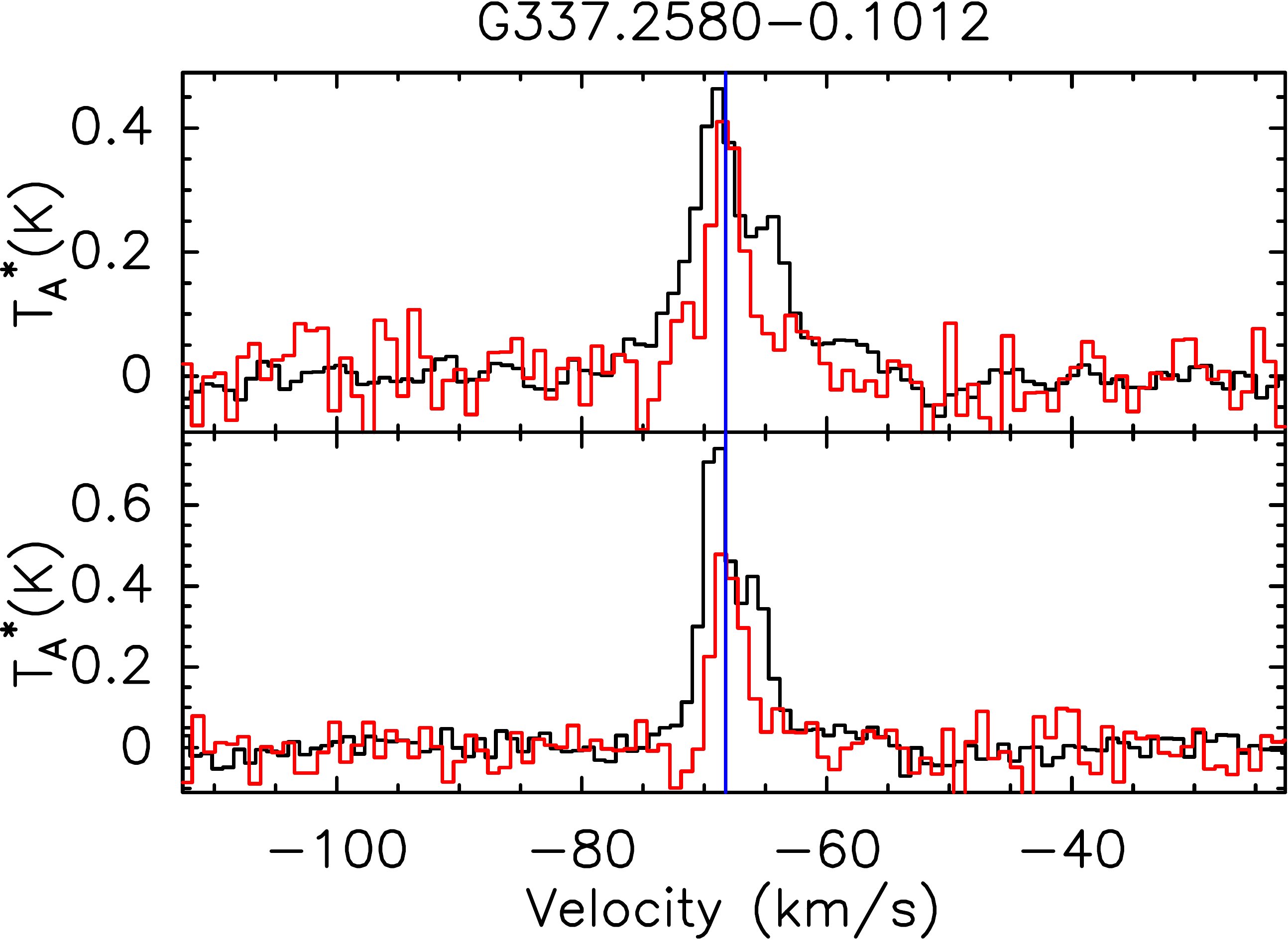}
      \caption{High-density tracers observed with Mopra. The source name is indicated above each panel. For each source, the HCO$^+(1-0)$ (black) and H$^{13}$CO$^+(1-0)$ (red) spectra are shown in the top sub-panel, whereas the HNC$(1-0)$ (black) and HN$^{13}$C$(1-0)$ (red) spectra are shown in the bottom sub-panel. The $\vlsr$ of the source is indicated in blue.}\label{fig:deep_spectra_online}
    \end{figure*}
    
    \begin{figure*}[tbp]
    \centering
      \includegraphics[width=0.9\columnwidth]{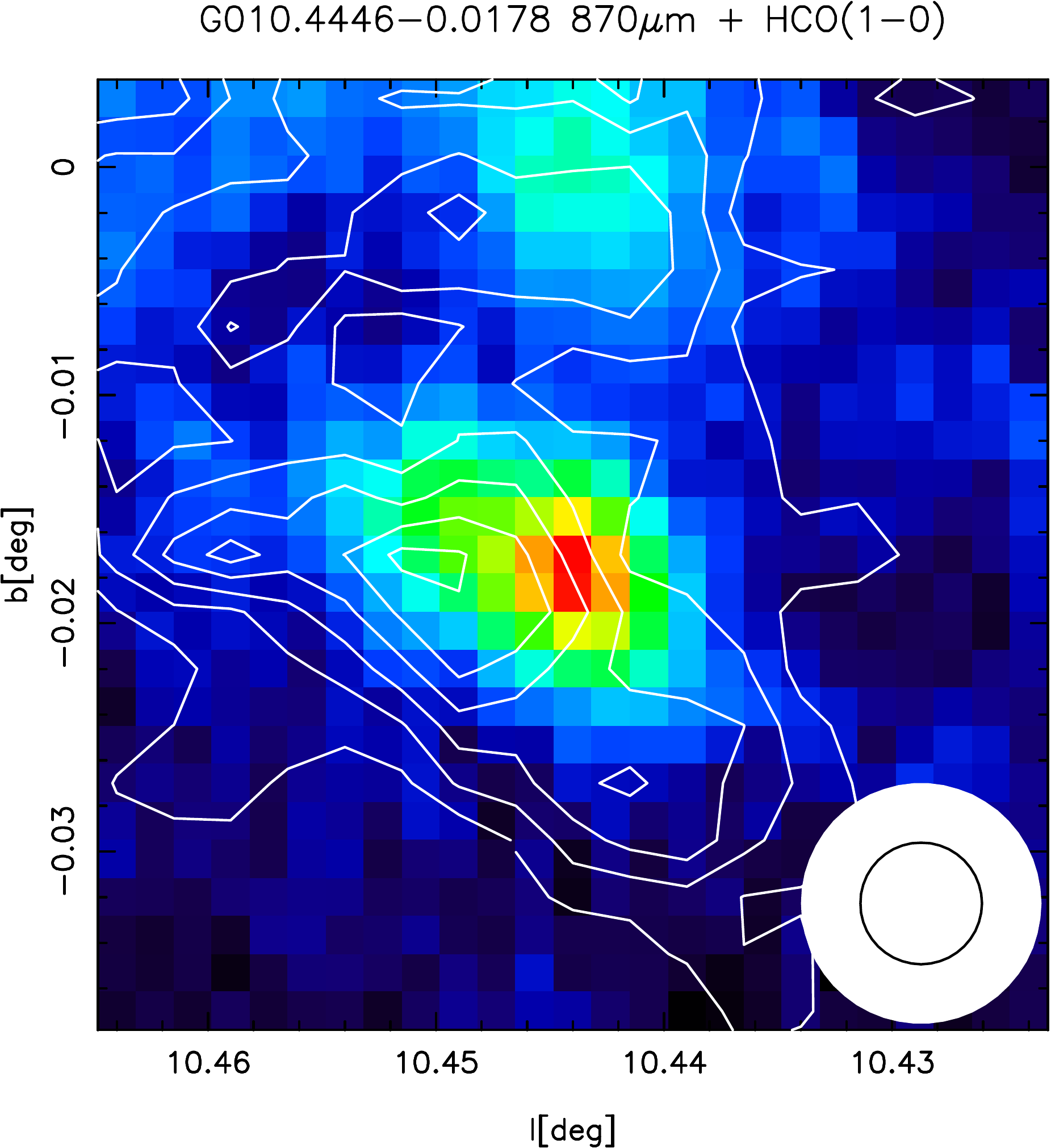}
      \hspace{0.07\columnwidth}
      \includegraphics[width=0.9\columnwidth]{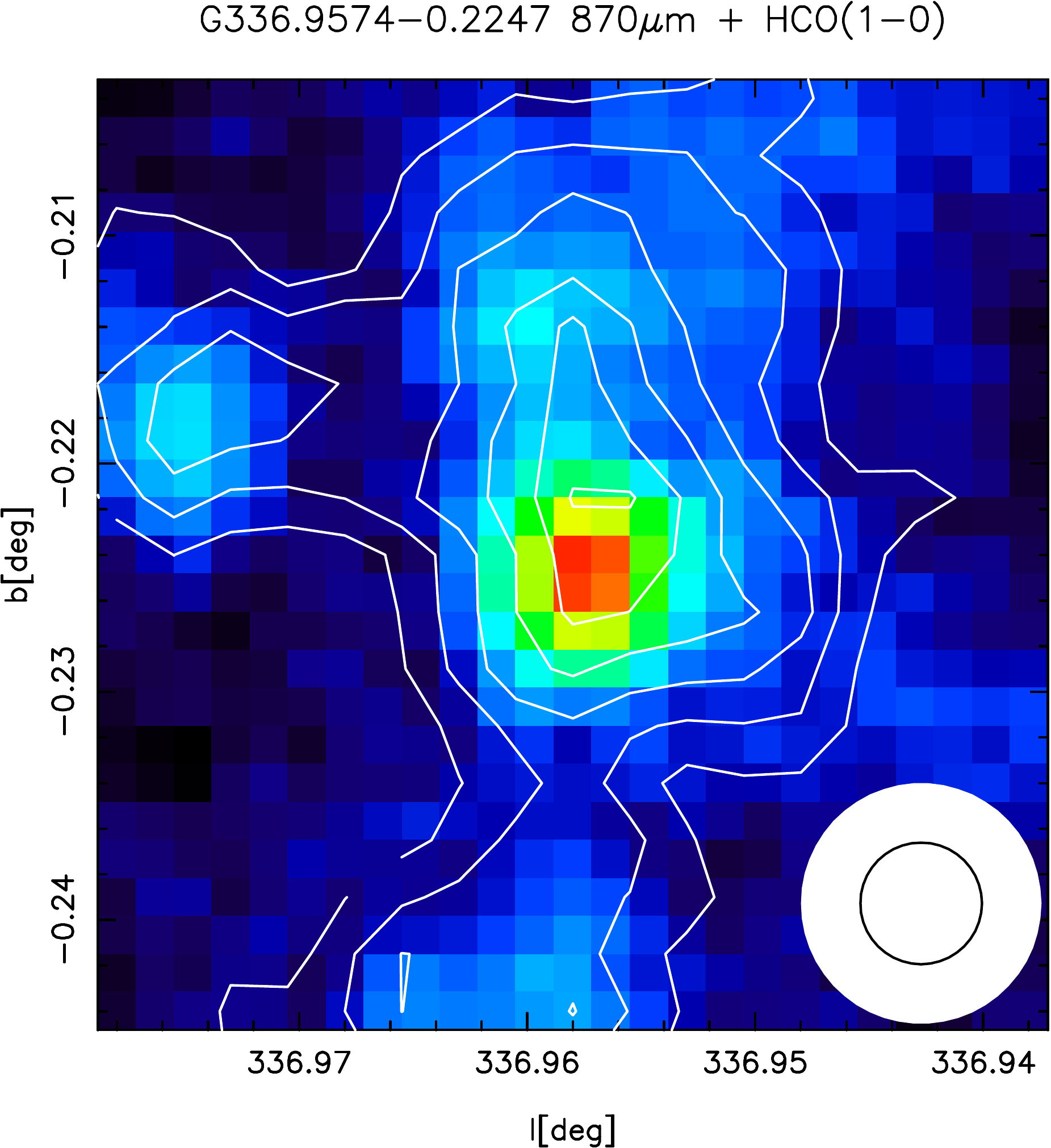}\\
      \vspace{1cm}
      \includegraphics[width=0.9\columnwidth]{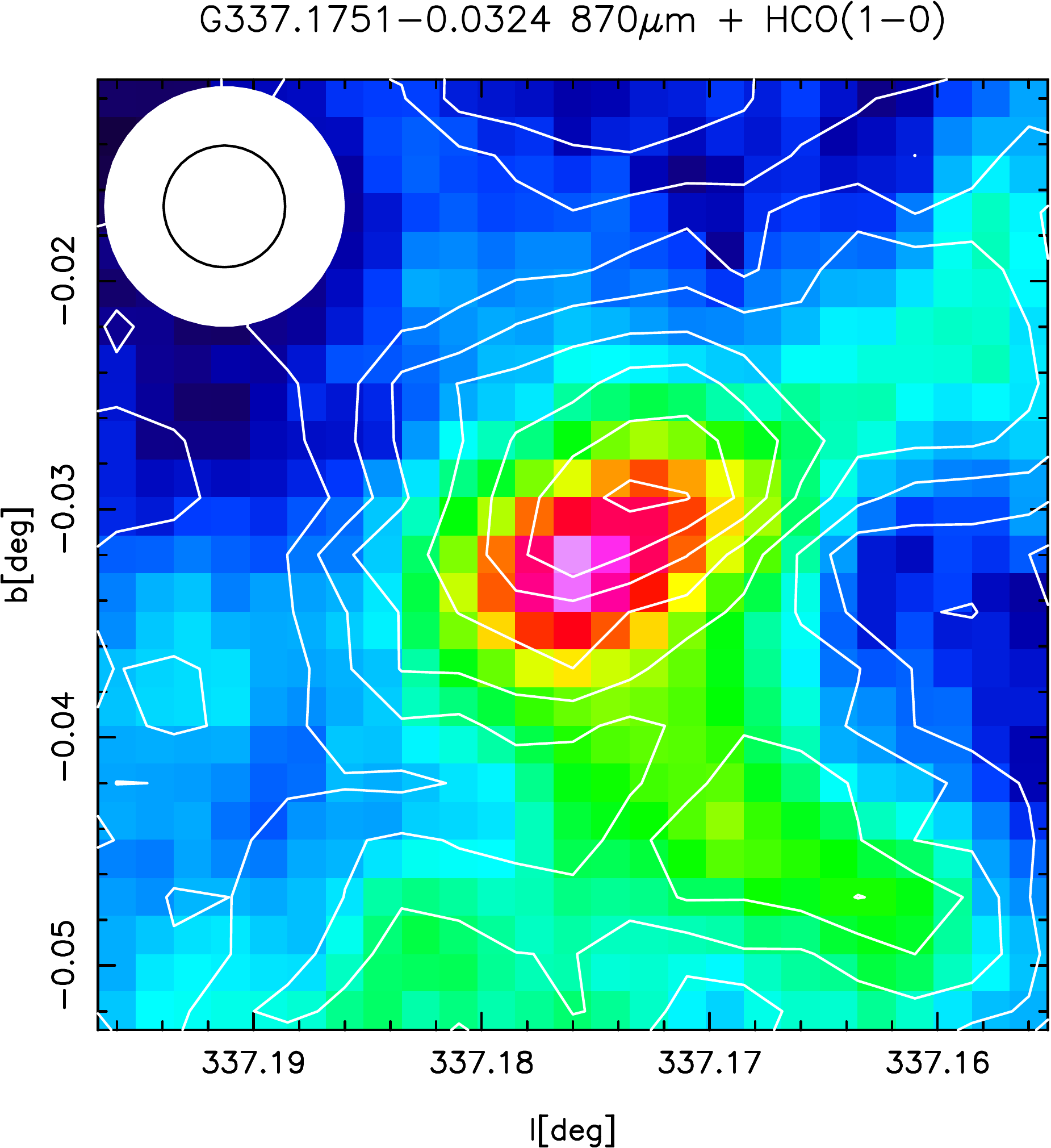}
      \hspace{0.07\columnwidth}
      \includegraphics[width=0.9\columnwidth]{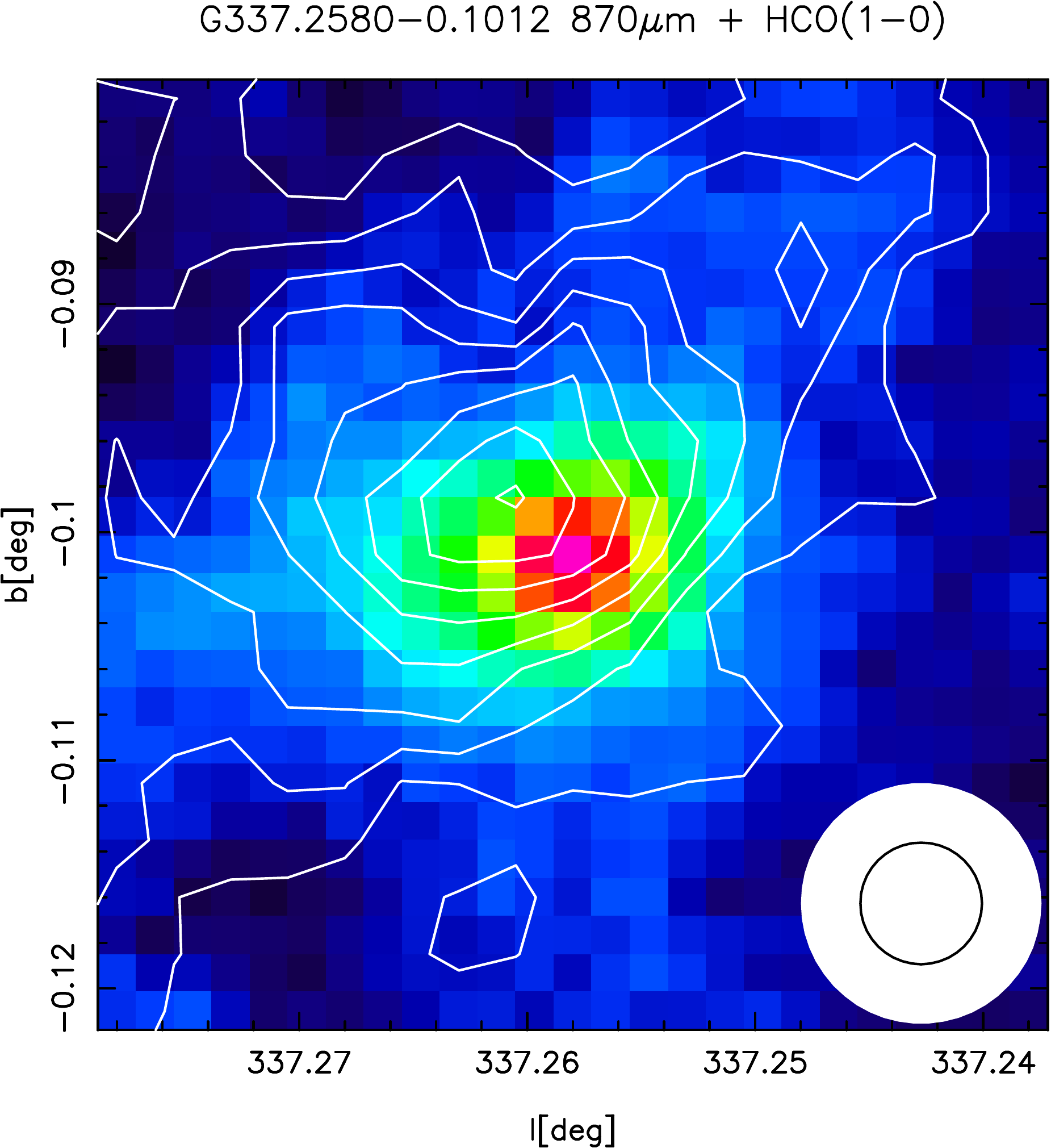}
      \caption{MALT90 HCO$^+(1-0)$ zeroeth moment map (contours) superimposed on the ATLASGAL sub-mm continuum emission at $870\mum$ (colourscale) for \firdconezero\ (top left), \firdcthreethreesix\ (top right), \firdcthreethreesevenpone\ (bottom left), and \firdcthreethreesevenptwo\ (bottom right). The Mopra and APEX beam sizes are indicated.}\label{fig:malt90_overlay}
    \end{figure*}

\end{appendix}

\end{document}